\documentclass[prd,aps,preprintnumbers,floats,floatfix,superscriptaddress,preprintnumbers,
showpacs,eqsecnum,nofootinbib,twocolumn]{revtex4-1}
\usepackage[utf8]{inputenc}
\usepackage{latexsym,array,theorem,mathrsfs,bm,float}
\usepackage{psfrag}
\usepackage{amsfonts,amsmath,amssymb,latexsym,array,afterpage,
theorem,mathrsfs,bm,float,epsfig,color,graphicx,tabularx,here,multirow}

\newcommand{\nn}{\nonumber \\}
\newcommand{\bea}{\begin{eqnarray}}
\newcommand{\ena}{\end{eqnarray}}
\newcommand{\beann}{\begin{eqnarray*}}
\newcommand{\enann}{\end{eqnarray*}}

\begin{document}
\baselineskip=12pt

\preprint{YITP-16-50} 
\preprint{IPMU16-0053} 

%<<<<<<<<<<<<< TITLE >>>>>>>>>>>>>>>%
%%
\title{Massive gravitons as dark matter and gravitational waves
}
%%
%<<<<<<<<<<<<< AUTHOR >>>>>>>>>>>>>>>%
%%
\author{Katsuki \sc{Aoki}}
\email{katsuki-a12@gravity.phys.waseda.ac.jp}
\affiliation{
Department of Physics, Waseda University,
Shinjuku, Tokyo 169-8555, Japan
}

\author{Shinji \sc{Mukohyama}}
\email{shinji.mukohyama@yukawa.kyoto-u.ac.jp}
\affiliation{Center for Gravitational Physics, Yukawa Institute for Theoretical Physics, Kyoto University, 606-8502, Kyoto, Japan}
\affiliation{Kavli Institute for the Physics and Mathematics of the
Universe (WPI), UTIAS, The University of Tokyo, Kashiwa, Chiba
277-8583, Japan}
%\affiliation{Kavli Institute for the Physics and Mathematics of the Universe, Todai Institutes for Advanced Study, University of Tokyo (WPI), 5-1-5 Kashiwanoha, Kashiwa, Chiba 277-8583, Japan}

%<<<<<<<<<<<<< DATE >>>>>>>>>>>>>>>%
\date{\today}

%======================================%
%<<<<<<<<<<<<< ABSTRACT >>>>>>>>>>>>>>>%
%======================================%
\begin{abstract}
We consider the possibility that the massive graviton is a viable candidate of dark matter in the context of bimetric gravity. We first derive the energy-momentum tensor of the massive graviton and show that it indeed behaves as that of dark matter fluid. We then discuss a production mechanism and the present abundance of massive gravitons as dark matter. Since the metric to which ordinary matter fields couple is a linear combination of the two mass eigenstates of bigravity, production of massive gravitons, i.e. the dark matter particles, is inevitably accompanied by generation of massless gravitons, i.e. the gravitational waves. Therefore, in this scenario some information about dark matter in our universe is encoded in gravitational waves. For instance, if LIGO detects gravitational waves generated by the preheating after inflation then the massive graviton with the mass of $\sim 0.01$ GeV is a candidate of the dark matter. 
\end{abstract}

%<<<<<<<<<<<<< PACS NUMBER >>>>>>>>>>>>>>>%
%\pacs{04.60.Cf, 04.50.Gh, 04.50.-h, 11.25.-w }

% 04.40.Dg	Relativistic stars: structure, stability, and oscillations (see also 97.60.-s Late stages of stellar evolution)
% 04.50.Gh : higher-dimensional Black holes
% 04.50.-h : Higher-dimensional gravity and other theories of gravity
% 04.60.Cf : gravitational aspects of String theory
% 11.25.-w : Strings and branes
% 04.50.Kd : Modified theories of gravity
% 98.80.-k : Cosmology
% 95.36.+x : Dark energy

\maketitle

%%

%%%%%%%%%%%%%%%%%%%%%%%%%%%%%%%%%%%%%%%%%%%%%%%%%%%%%%%%%%%%%%%
%%%%%%%%%%%%%%%%%%%%%%%%%%%%%%%%%%%%%%%%%%%%%%%%%%%%%%%%%%%%%%%
%%%%%%%%%%%%%%%%%%%%%%%%%%%%%%%%%%%%%%%%%%%%%%%%%%%%%%%%%%%%%%%
%%%%%%%%%%%%%%%%%%%%%%%%%%%%%%%%%%%%%%%%%%%%%%%%%%%%%%%%%%%%%%%

\section{Introduction}

The massive graviton has long received much attentions from both theoretical and phenomenological aspects, ever since the linear theory of the massive graviton was constructed by Fierz and Pauli in 1939~\cite{Fierz:1939ix}. Although generic nonlinear extensions of the Fierz-Pauli theory lead to an unstable degree of freedom called Boulware-Deser ghost\cite{Boulware:1973my}, the nonlinear ghost-free massive gravity was constructed by de Rham {\it et al.} in 2010~\cite{deRham:2010ik,deRham:2010kj}. Furthermore, the nonlinear ghost-free massive gravity is generalized to the bigravity theory \cite{Hassan:2011zd} and the multigravity theory \cite{Hinterbichler:2012cn}
(See \cite{Hinterbichler:2011tt,deRham:2014zqa,Schmidt-May:2015vnx} for reviews). 
In this paper, we assume the bigravity theory which contains a massive graviton as well as a massless graviton.

If the massive graviton exists, 
the gravity would be modified around the scales of the Compton wavelength of the massive graviton.
This modification of gravity yields various phenomenological features depending on the graviton mass 
(see \cite{Will:2014kxa,Murata:2014nra,TheLIGOScientific:2016src} for experimental constraints on the graviton mass).
Many studies addressed to explain the present accelerating expansion of the Universe by the tiny graviton mass as $m\sim 10^{-33}$ eV
\cite{Volkov:2011an,vonStrauss:2011mq,Berg:2012kn,Comelli:2011zm,Maeda:2013bha,Akrami:2012vf,Akrami:2013ffa,Aoki:2013joa}.
Another possibility is to explain the origin of dark matter when the graviton mass is $m\gtrsim 10^{-27}$ eV.
When a matter field is introduced in the ``dark'' sector, it acts as the dark matter in the physical sector
through the gravity interaction \cite{Aoki:2013joa,Aoki:2014cla}.

In the present paper, however, we focus on a particle aspect of the massive graviton.
In general relativity (GR),
while the graviton is the mediator of the gravity,
the graviton itself is a source of the gravitational field,
whose energy-momentum tensor was derived by Isaacson \cite{Isaacson:1968zza}.
Hence one expects that the massive graviton is also a gravitational source
in the bigravity theory.
In particular, if the massive graviton behaves like just a massive field,
the massive graviton itself is a candidate of the dark matter.
Indeed, by calculating the energy-momentum tensor of the massive graviton,
we present that the massive graviton is a gravitational source
and it acts as a dark matter in the bigravity theory.

Since the bigravity contains both massless and massive gravitons,
when the massive graviton is generated,
the massless graviton is also generated.
The massless gravitons would then be observed as a gravitational wave background.
Therefore, if the massive graviton in bigravity is dark matter,
the gravitational wave background can carry information about the dark matter.
As an example, we assume a production of the massive graviton from the preheating. 
The gravitational waves from the preheating have been discussed in 
\cite{Khlebnikov:1997di,Khlebnikov:1997di,Easther:2006gt,Easther:2006vd,Felder:2006cc,Dufaux:2007pt,GarciaBellido:2007af,GarciaBellido:2007dg,Dufaux:2008dn,Price:2008hq}.
In the bigravity, massive gravitons are also generated from the preheating.
We find that, if the massive graviton is the dominant component of the dark matter,
the graviton mass can be estimated by observations of the gravitational wave background.
In particular, if LIGO and Virgo detectors observe the gravitational wave background originated from the preheating,
the massive graviton with $m\sim 0.01$ GeV is a viable candidate of the dark matter.

The paper is organized as follows.
The nonlinear ghost-free bigravity theory is introduced in Sec. \ref{sec_bigravity}.
In Sec. \ref{MGDM}, we derive the energy-momentum tensor of the massive graviton,
and confirm that the massive graviton can be a dark matter.
We discuss the generation of the massive graviton from the preheating 
and observational implications of the massive graviton dark matter in Sec. \ref{sec_abundance}.
We summarize our results and give some remarks in Sec. \ref{summary}.
In Appendix \ref{appendix}, 
we detail the definition and the derivation of the energy-momentum tensor of the massive graviton.

\section{Bigravity theory}
\label{sec_bigravity}
The nonlinear ghost-free bigravity action \cite{Hassan:2011zd} is given by
\begin{align}
 S &=\frac{1}{2 \kappa _g^2} \int d^4x \sqrt{-g}R(g)+ \frac{1}{2 \kappa _f^2}
 \int d^4x \sqrt{-f} \mathcal{R}(f) 
 \nn
&+\frac{m^2}{ \kappa ^2} \int d^4x \sqrt{-g} \mathscr{U}(g,f) 
+S_{\rm m}\,,
\label{action}
\end{align}
where $g_{\mu\nu}$ and $f_{\mu\nu}$ are two dynamical metrics, and
$R(g)$ and $\mathcal{R}(f)$ are their Ricci scalars.
The parameters $\kappa_g^2$ and $\kappa_f^2$ are 
the corresponding gravitational constants, 
while $\kappa$ is defined by $\kappa^2=\kappa_g^2+\kappa_f^2$.
To admit the Minkowski spacetime as a vacuum solution,
we restrict the potential $\mathscr{U}$ as the form
\begin{align}
\mathscr{U}&=\mathscr{U}_2(\mathcal{K})+c_3\mathscr{U}_3(\mathcal{K})+c_4\mathscr{U}_4(\mathcal{K})\,,
\label{dRGT_potential}
\\
\mathscr{U}_2(\mathcal{K})&=-\frac{1}{4}\epsilon_{\mu\nu\rho\sigma} 
\epsilon^{\alpha\beta\rho\sigma}
{\mathcal{K}^{\mu}}_{\alpha}{\mathcal{K} ^{\nu}}_{\beta}\,, \nn
\mathscr{U}_3(\mathcal{K})&=-\frac{1}{3!}\epsilon_{\mu\nu\rho\sigma} 
\epsilon^{\alpha\beta\gamma\sigma}
{\mathcal{K} ^{\mu}}_{\alpha}{\mathcal{K} ^{\nu}}_{\beta}{\mathcal{K} ^{\rho}}_{\gamma}\,, 
\\
\mathscr{U}_4(\mathcal{K})&=-\frac{1}{4!}\epsilon_{\mu\nu\rho\sigma} 
\epsilon^{\alpha\beta\gamma\delta}
{\mathcal{K} ^{\mu}}_{\alpha}{\mathcal{K} ^{\nu}}_{\beta}{\mathcal{K} ^{\rho}}_{\gamma}
{\mathcal{K}^{\sigma}}_{\delta}\,,
\nonumber
\end{align}
with
\begin{align}
\mathcal{K}^{\mu}{}_{\nu}=\delta^{\mu}{}_{\nu}
-\left(\sqrt{g^{-1}f}\right)^{\mu}{}_{\nu}\,,
\end{align}
where $\left(\sqrt{g^{-1}f}\right)^{\mu}{}_{\nu}$ is defined by the relation
\begin{align}
\left(\sqrt{g^{-1}f}\right)^{\mu}{}_{\rho}
\left(\sqrt{g^{-1}f}\right)^{\rho}{}_{\nu}
=f^{\mu\rho}g_{\rho\nu}\,.
\end{align}
Then $g_{\mu\nu}=f_{\mu\nu}=\eta_{\mu\nu}$ is a vacuum solution of the bigravity,
and the parameter $m$ describes the mass of
the massive graviton propagating on the Minkowski background.

We define perturbations of the two metrics as
\begin{align}
\delta g{}_{\mu\nu}&:=g_{\mu\nu}-\eta_{\mu\nu}\,, \nn
\delta f{}_{\mu\nu}&:=f_{\mu\nu}-\eta_{\mu\nu}\,.
\label{perturbations}
\end{align}
Note that either $\delta g_{\mu\nu}$ or $\delta f_{\mu\nu}$ is not mass eigenstate.
At the linear order of the perturbations, the mass eigenstates are defied by
\begin{align}
h_{\mu\nu}&:= \frac{\kappa_f}{\kappa_g \kappa}\delta g{}_{\mu\nu} 
+\frac{\kappa_g}{\kappa_f \kappa}\delta f{}_{\mu\nu}
\,, 
\label{massless_eigenstate}\\
\varphi_{\mu\nu}&:=\frac{1}{\kappa}\left( \delta g{}_{\mu\nu}-\delta f{}_{\mu\nu} \right)
\,,
\label{massive_eigenstate}
\end{align}
where $h_{\mu\nu}$ and $\varphi_{\mu\nu}$ are the massless and the massive eigenstate
with mass dimension one, respectively.
A nonlinear extension of mass eigenstates was discussed in \cite{Hassan:2012wr}.

The matter action $S_{\rm m}$ can be divided into three types:
\begin{align}
S_{\rm m}=S_g(g,\psi_g)+S_f(f,\psi_f)+S_d(g,f,\psi_d)
\,,
\end{align}
where first two types of matter fields couple to either $g_{\mu\nu}$ or $f_{\mu\nu}$, while the third type couples to both metrics. The matter fields that couple to only one metric do not spoil the structure of the gravitational part of the theory that eliminates the (would-be) BD ghost. On the other hand, matter fields that couple to both metrics generically reintroduce the BD ghost~\cite{Yamashita:2014fga,deRham:2014fha,deRham:2014naa,Heisenberg:2014rka,Hinterbichler:2015yaa,Heisenberg:2015iqa,deRham:2015cha}. This would imply that the matter should couple to only one metric. One way to avoid the difficulty of the double matter coupling was recently proposed in the context of the partially constrained vielbein formulation that breaks Lorentz invariance at the cosmological scale~\cite{DeFelice:2015yha}, making it possible to couple matter fields simultaneously to both metrics without the BD ghost at all scales. However, in the present paper, for simplicity we shall not consider the double matter coupling. Furthermore, we simplify the system by restricting our considerations to the first type of matter fields only, i.e. those that couple to $g_{\mu\nu}$ only. Even in this simplest setup, since the mass eigenstates of the gravitons are defined by \eqref{massless_eigenstate} and \eqref{massive_eigenstate}, the matter couples to both massless and massive gravitons, simultaneously.

In this paper, as already stated above, we assume that all matter fields couple minimally to $g_{\mu\nu}$~\footnote{If we introduce a matter field coupling to $f_{\mu\nu}$, the matter can act as a dark matter component in $g_{\mu\nu}$ \cite{Aoki:2013joa, Aoki:2014cla}. However, in this paper, we discuss whether the massive graviton itself can be a candidate of dark matter or not, thus we do not consider such a matter field.}.
The quadratic action is expressed as
\begin{align}
S_2&=\int d^4x \Biggl[ 
\frac{1}{\kappa_g^2}\mathcal{L}_{\rm EH}\big[\delta g \big]+
\frac{1}{\kappa_f^2}\mathcal{L}_{\rm EH} \big[\delta f \big]
\nn
&\qquad \qquad\quad
+\mathcal{L}_{\rm FP} \big[\varphi \big] 
+\frac{1}{2}\delta g{}_{\mu\nu}T^{\mu\nu}_{\rm m}
 \Biggl] \nn
&=\int d^4x \left[ \mathcal{L}_{\rm EH}\big[h \big]+\frac{1}{2M_{\rm pl}}h_{\mu\nu}T^{\mu\nu}_{\rm m} \right]
\nn
&+\int d^4x \left[\mathcal{L}_{\rm EH}\big[\varphi \big]
+\mathcal{L}_{\rm FP}\big[\varphi \big]+\frac{1}{2M_G}\varphi_{\mu\nu}T^{\mu\nu}_{\rm m} \right]\,,
\label{quad-act}
\end{align}
where $T_{\rm m}^{\mu\nu}$ is the matter energy-momentum tensor. 
The quadratic Einstein-Hilbert Lagrangian and the Fierz-Pauli mass term
for a symmetric tensor field $\chi_{\mu\nu}$ are defined by
\begin{align}
\mathcal{L}_{\rm EH}[\chi  ]&=
-\frac{1}{4}\chi^{\mu\nu} \mathcal{E}_{\mu\nu,\alpha\beta}\chi^{\alpha\beta}
\, , 
\label{EH_quadratic}\\
\mathcal{L}_{\rm FP}[\chi ]&=-\frac{m^2}{8}\left( \chi_{\mu\nu}\chi^{\mu\nu}-\chi^2\right) \,,
\label{FP_quadratic}
\end{align}
where
\begin{align}
\mathcal{E}_{\mu\nu,\alpha\beta}\chi^{\alpha\beta}
&=-\frac{1}{2}\partial^2 \chi_{\mu\nu}-\frac{1}{2}\partial_{\mu}\partial_{\nu} \chi
+\partial_{\alpha}\partial_{(\nu}\chi^{\alpha}{}_{\mu)}
\nn
&+\frac{1}{2}\eta_{\mu\nu}\left( \partial^2 \chi-\partial_{\alpha}\partial_{\beta}\chi^{\alpha\beta}\right)
\,,
\end{align}
with the notation $\chi=\chi^{\mu}{}_{\mu}$.
The gravitational coupling constants are defined by
\begin{align}
M_{\rm pl}:=\frac{\kappa}{\kappa_g\kappa_f}\,, \quad M_G:=\frac{\kappa}{\kappa_g^2} = \frac{\kappa_f}{\kappa_g}M_{\rm Pl} \,.
\end{align}

\begin{figure}[htbp]
 \begin{center}
  \includegraphics[width=70mm]{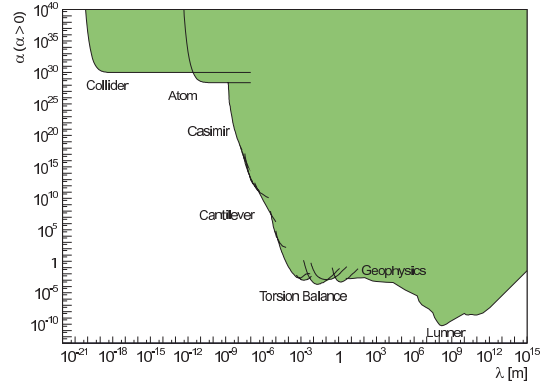}
 \end{center}
 \caption{Experimental constraints on the gravitational potential \eqref{gravitational_potential} adapted from Ref. \cite{Murata:2014nra}.
The colored region is the excluded area at $95\%$ confidence level (see \cite{Murata:2014nra} and references therein for details).}
 \label{alpha-lambda-fullscale}
\end{figure}

Since the matter couples to $g_{\mu\nu}$, 
the gravitational potential observed by the physical matter is defined by $\Phi:=-\delta g{}_{00}/2$.
Using the weak field approximation, one can obtain
\begin{align}
\Phi=-\frac{GM}{r}\left(1+\alpha e^{-r/\lambda} \right)
\,, \label{gravitational_potential}
\end{align}
where $G:=1/8\pi M_{\rm pl}^2, \alpha:=\frac{4}{3}M_{\rm pl}^2/M_G^2,\lambda=m^{-1}$.
The experimental constraints on the gravitational potential \eqref{gravitational_potential}
is summarized in Fig. \ref{alpha-lambda-fullscale} from \cite{Murata:2014nra}.
Note that this constraint does not include the effect of Vainshtein mechanism \cite{Vainshtein:1972sx}.
The linear approximation is no longer valid inside the Vainshtein radius\cite{Babichev:2013pfa, Enander:2015kda, Aoki:2016eov}.
Hence, the constraints on the large scales  are subject to discussion.

%%%%%%%%%%%%%%%%%%%%%%%%%%%%%%%%%%%%%%%%%%%%%%%%%%%%%
%%%%%%%%%%%%%%%%%%%%%%%%%%%%%%%%%%%%%%%%%%%%%%%%%%%%%
%%%%%%%%%%%%%%%%%%%%%%%%%%%%%%%%%%%%%%%%%%%%%%%%%%%%%
%%%%%%%%%%%%%%%%%%%%%%%%%%%%%%%%%%%%%%%%%%%%%%%%%%%%%
%%%%%%%%%%%%%%%%%%%%%%%%%%%%%%%%%%%%%%%%%%%%%%%%%%%%%
%%%%%%%%%%%%%%%%%%%%%%%%%%%%%%%%%%%%%%%%%%%%%%%%%%%%%

\section{Massive graviton as dark matter}
\label{MGDM}

In this section, we discuss whether the massive graviton  can be a dark matter or not. We focus on scales well inside the cosmological horizon but well outside the Vainshtein radius. Hence we can analyze the system based on a perturbative approach around the Minkowski background. The energy-momentum tensor of the massive graviton is evaluated in a way similar to the standard case of GR~\cite{Isaacson:1968zza}.

First we discuss the free propagating massless and massive gravitational waves.
In vacuum, the equations of motion at linear order are given by
\begin{align}
\mathcal{E}_{\mu\nu,\alpha\beta}h^{\alpha\beta}=0\,,
\\
\mathcal{E}_{\mu\nu,\alpha\beta}\varphi^{\alpha\beta} +\frac{m^2}{2}(\varphi_{\mu\nu}-\varphi \eta_{\mu\nu})=0\,.
\end{align}
Since the massless graviton has a gauge symmetry, we can chose the transverse-traceless gauge for the massless eigenstate, i.e., 
\begin{align}
\partial_{\mu}h^{\mu}{}_{\nu}=0\,, \quad
h=0\,, \quad
h_{\mu\nu}u^{\nu}=0\,,
\end{align}
where $u^{\mu}$ is a timelike vector. Since the massive graviton does not enjoy the gauge symmetry, so we cannot impose any gauge condition for the massive graviton. However, in vacuum, we can obtain the transverse-traceless condition from the equation of motion:
\begin{align}
\partial_{\mu}\varphi^{\mu}{}_{\nu}=0\,, \quad
\varphi=0\,.
\label{TT_condition}
\end{align}
As a result, the equations of motion are expressed as
\begin{align}
\partial^2 h_{\mu\nu}&=0\,,
\label{EH_eom}
\\
(\partial^2-m^2)\varphi_{\mu\nu}&=0
\,.
\label{FP_eom}
\end{align}
The equations are the Klein-Gordon equations with and without the mass term, thus we can easily find their solutions. However, the explicit forms of the solutions are not necessary to evaluate the energy-momentum tensor.

As is well known in GR, the division of the spacetime geometry into a background and gravitational waves requires a separation of scales for the two: the length or/and time scale associated with the perturbation should be sufficiently shorter than the scale associated with the smooth background~\cite{Isaacson:1968zza}. In this situation the energy-momentum tensor of gravitational waves is defined by the second order part of the perturbed Einstein equation averaged over a length or/and time scale between the two scales. The same assumption and procedure can be employed to define the energy-momentum tensor of the massless graviton in the context of bigravity. Specifically, the assumption of a large hierarchy of scales makes it possible for us to perform integration by part, e.g. as
\begin{align}
\langle \partial_{\rho} h_{\mu\nu} h_{\alpha\beta} \rangle
 \approx 
-\langle  h_{\mu\nu} \partial_{\rho} h_{\alpha\beta} \rangle
\,,
\end{align}
similarly to the standard procedure in GR \cite{Isaacson:1968zza}, where the symbol $\langle \cdots \rangle$ denotes an average over a spacetime region with a size larger than the corresponding scale of the perturbation but smaller than the scale of the background,
defined through an appropriate window function.

For the massless graviton with the transverse-traceless gauge (\ref{TT_condition}), in both GR and bigravity, the integration by part can be applied to the time derivative as well as the spatial derivatives even if the average is over a spatial region, provided that the gravitational wave over the region of integration can be considered as a wave propagating to one direction. For example, in a region sufficiently far from a finite-size source a solution to the wave equation propagating to, say, the $z$ direction is written as $F(t-z)$ and thus $\partial_t$ applied to it can be replaced by $-\partial_z$ before performing the spatial integration by part and then $\partial_z$ acted on another function of the form $G(t-z)$ can be replaced by $-\partial_t$. On the other hand, this argument does not apply to the massive graviton since a wave of a massive field changes its shape as it propagates in one direction. Moreover, even for the massless graviton, in either GR or bigravity, this argument does not seem to be valid for stochastic gravitational waves, which come from every direction to every point. In the present paper, we thus employ an average over a spacetime region to make it possible to do integration by part.

In order to define the stress-energy tensor of the massive graviton in bigravity, we thus assume that the length and time scales associated with the background are sufficiently longer than the corresponding scales of the massive graviton mode, at least in the spacetime region where we are about to evaluate the stress-energy tensor. We are then able to define the stress-energy tensor of the massive graviton by averaging the contribution of the massive graviton to the second order part of the perturbed Einstein equation over a spacetime region whose size is greater than the Compton wavelength but shorter than the scales of the background. In this case we are able to perform integration by part, e.g. as
\begin{align}
\langle \partial_{\rho} \varphi_{\mu\nu} \varphi_{\alpha\beta} \rangle
 \approx 
-\langle  \varphi_{\mu\nu} \partial_{\rho} \varphi_{\alpha\beta} \rangle
\,,
\end{align}
where the symbol $\langle \cdots \rangle$ denotes an average over a spacetime region. The explicit calculation of the stress-energy tensors for the massless and massive gravitons in bigravity is summarized in Appendix \ref{appendix}.

We further demand that the length scale associated with the smooth background is longer than the Compton wavelength of the massive graviton mode. In this situation, gravity is basically mediated by the massless graviton: while matter fields propagate on the metric $g_{\mu\nu}$ and its perturbation is a linear combination of the massless and massive graviton modes, the latter mode is exponentially suppressed at the length scale of the background. Hence, only the Einstein equation of the massless graviton is relevant. Including the energy-momentum tensors of massless and massive gravitons, the equation of motion for the massless graviton, after averaging over a spacetime region with the size larger than the scales of the perturbation but smaller than the scales of the background, is given by
\begin{align}
\mathcal{E}^{\mu\nu,\alpha\beta}h_{\alpha\beta}
=\frac{1}{M_{\rm pl}}(T_{\rm m}^{\mu\nu}+T_{\rm gw}^{\mu\nu}+T_G^{\mu\nu})
\,,
\end{align}
where 
$T_{\rm gw}^{\mu\nu}$ is the usual energy-momentum tensor of the massless graviton,
while $T_G^{\mu\nu}$ is the energy-momentum tensor of the massive graviton. As shown in Appendix \ref{appendix}, they are given by 
\begin{align}
T_{\rm gw}^{\mu\nu}&=\frac{1}{4}
\langle h^{\alpha\beta,\mu} h_{\alpha\beta}{}^{,\nu} \rangle
\,,\\
T_G^{\mu\nu}&=
\frac{1}{4} \langle \varphi^{\alpha\beta,\mu} \varphi_{\alpha\beta}{}^{,\nu} \rangle
\,,
\end{align}
where $_{,\mu}$ denotes a partial derivative~\footnote{Rigorously speaking, $T_{\rm gw}^{\mu\nu}$ and $T_G^{\mu\nu}$ must be called ``pseudotensors''. In the present paper, for simplicity we shall call them tensors.}. These energy-momentum tensors are also obtained from the Noether's theorem (see Appendix \ref{appendix}).

When the massive graviton is non-relativistic, the massive graviton indeed behaves like a dust as a source of the massless graviton. At the rest frame of the massive graviton, the energy-momentum tensor is indeed given by
\begin{align} 
T^{\mu\nu}_G= \frac{m^2}{4}{\rm diag}[\langle \varphi^{\alpha\beta} \varphi_{\alpha\beta} \rangle,0,0,0]
\,.
\end{align}

If the massive graviton is the dark matter, the massive gravitons have to survive until today. However, since the graviton couples universally to matter fields, the massive graviton can decay to light particles. The total decay rate of massive graviton \cite{Han:1998sg,Giudice:1998ck,Hewett:1998sn} is given by 
\begin{align}
\Gamma_G\sim 0.1 \frac{m^3}{M_G^2}\,.
\end{align}
If the decay rate of massive graviton is larger than the present Hubble parameter, the massive graviton cannot be relict at present. By demanding that the decay rate be lower than the present Hubble parameter, an upper bound on the graviton mass is thus given by 
\begin{align}
m \lesssim 0.01 \left(\frac{M_G}{M_{\rm pl}} \right)^{2/3}
\; {\rm GeV}\,.
\end{align}
On the other hand, the existence of dark matter in galaxies gives a lower bound on the graviton mass. Since the massive graviton should be confined in galaxies, the de Broglie wavelength of the massive graviton $2\pi/(mv)$ should be smaller than kpc scale. Using a typical velocity $v\sim 10^{-3}$ in the halo, a lower bound of the graviton mass is given by 
\begin{align}
 m \gtrsim 10^{-23} \; {\rm eV}\,. 
\end{align}
In summary, when the mass is in the range
\begin{align}
10^{-23}\; {\rm eV} \lesssim m \lesssim 
0.01 \left(\frac{M_G}{M_{\rm pl}} \right)^{2/3}
\; {\rm GeV}\,,
\label{constraint_for_DM}
\end{align}
the massive graviton can be a candidate of dark matter.

%%%%%%%%%%%%%%%%%%%%%%%%%%%%%%%%%%%%%%%%%%%%%%%%%%%%%
%%%%%%%%%%%%%%%%%%%%%%%%%%%%%%%%%%%%%%%%%%%%%%%%%%%%%
%%%%%%%%%%%%%%%%%%%%%%%%%%%%%%%%%%%%%%%%%%%%%%%%%%%%%
%%%%%%%%%%%%%%%%%%%%%%%%%%%%%%%%%%%%%%%%%%%%%%%%%%%%%
%%%%%%%%%%%%%%%%%%%%%%%%%%%%%%%%%%%%%%%%%%%%%%%%%%%%%
%%%%%%%%%%%%%%%%%%%%%%%%%%%%%%%%%%%%%%%%%%%%%%%%%%%%%

\section{Present abundance of massive graviton}
\label{sec_abundance}

One of the simplest scenarios of the generation of the massive graviton in the early universe would be through inflation as discussed in \cite{Dubovsky:2004ud,Pshirkov:2008nr}. In this case, however, the Hubble expansion rate during inflation must be larger than the graviton mass to produce sufficient amount of massive gravitons for dark matter. In our present setup of bigravity, this would imply that the Higuchi bound tends to be violated and thus there appears a ghost at the linear level~\cite{Higuchi:1986py,Higuchi:1989gz,Grisa:2009yy,Comelli:2012db,Comelli:2014bqa,DeFelice:2014nja}. This would at least invalidate the perturbative approach~\cite{Aoki:2015xqa} and thus we shall not consider generation of massive graviton during inflation in the present paper.

Instead of the production by inflation, we thus consider generation of the massive graviton through the preheating after inflation. During preheating, the inflaton decays to inhomogeneous modes of itself and/or some other fields and then large inhomogeneities can be created. This kind of field bubble is a classical source of gravitational waves. The peak momentum $k_*=|{\mathbf k}_*|$ and the energy density $\rho_{\rm gw}^*$ of the generated massless gravitational wave are roughly estimated as
\begin{align}
k_* \sim 1/R_*\,,\quad
\rho^*_{\rm gw}\sim \, \alpha\, (R_* H_*)^2 \rho_*
\end{align}
where $R_*,H_*$ and $\rho_*$ are the typical size of the field bubble, the Hubble expansion rate, and the energy density at the time of production, respectively, and we have included a numerical factor $\alpha$ that varies from one model to another ($\alpha\simeq 0.1$ for chaotic inflation, for example)~\cite{Felder:2006cc,Dufaux:2007pt,Dufaux:2008dn}. The typical size $R_*$ and the numerical factor $\alpha$ can be evaluated when we assume a concrete preheating model. In the present paper, however, we take a phenomenological attitude and treat $R_*$ and $\alpha$ as a free parameter to discuss a model independent prediction. The present frequency and the density parameter of the gravitational wave background are then given by
\begin{align}
f \sim \frac{4 \times 10^{10}}{R_* \rho_*^{1/4}} {\rm Hz}
\,,
\quad
h^2\Omega_{\rm gw} \sim 10^{-5}\,\alpha\, (R_* H_*)^2
\,.
\label{GW_preheating}
\end{align}

Note that in this model, the gravitational waves are created at the sub-horizon scale which remain the sub-Horizon scale until today. We can assume the graviton mass is larger than the Hubble expansion rate at the time of production of gravitational waves so that the Higuchi instability is avoided. Therefore, the cosmic history of the amplitude of gravitational waves can be discussed by using the linear theory until today.

In the sub-horizon scale, we can ignore the effect of the expansion of the Universe to discuss the generations of the massless and the massive gravitons. Hence we can use the equations on the Minkowski background. For the massless graviton, the equation of motion with a source is expressed by
\begin{align}
\partial^2 h^{\mu\nu}&=-\frac{2}{M_{\rm pl}}
S^{\mu\nu}
\,, 
\label{massless_KG_with_s}
\end{align}
where $S^{\mu\nu}$ will be specified in \eqref{def_S} below and we have chosen the harmonic gauge
\begin{align}
\partial_{\mu}h^{\mu}{}_{\nu}=\frac{1}{2}\partial_{\nu}h
\,.
\end{align}
On the other hand, the equation of motion for the massive graviton is given by
\begin{align}
(\partial^2-m^2)\varphi^{\mu\nu}=
-\frac{2}{M_G}J^{\mu\nu}
\,,
\label{massive_KG_with_s}
\end{align}
where the massive graviton must satisfy the constraint equations
\begin{align}
\partial_{\mu}\varphi^{\mu\nu}=\partial^{\nu}\varphi
\,,\quad
\frac{m^2}{2}\varphi=-\frac{1}{3M_G}T_{\rm m}
\,.
\end{align}
The source terms for massless and massive gravitons are given by
\begin{align}
S^{\mu\nu}&:=  T_{\rm m}^{\mu\nu}-\frac{1}{2}\eta^{\mu\nu} T_{\rm m} 
\,,
\label{def_S}\\
J^{\mu\nu}&:=
T_{\rm m}^{\mu\nu}
-\frac{1}{3}\left(\eta^{\mu\nu}-\frac{\partial^{\mu}\partial^{\nu}}{m^2}\right)
T_{\rm m}\,.\label{def_J}
\end{align}

Using the retarded Green's function
\begin{align}
&\qquad G_R(x-y;p)
\nn
&=\theta(x^0-y^0) 
\int \frac{d^3{\mathbf p}}{(2\pi)^3} \frac{-i}{2p^0}
\left(e^{ip(x-y)}-e^{-ip(x-y)}\right)
,
\end{align}
the solutions of Eqs. \eqref{massless_KG_with_s} and \eqref{massive_KG_with_s} can be constructed. We denote $k^{\mu}$ as the four-momentum of the massless graviton and $p^{\mu}$ as the four-momentum of the massive graviton with $p^{\mu}p_{\mu}=-m^2$. We evaluate the solutions after the source vanishes, i.e., after the preheating. Choosing the coordinate $u^{\mu}=\delta^{\mu}_0$ in the transverse-traceless gauge, the solutions are given by
\begin{align}
h_{0 \mu }(x)&=0\,, \nn
h_{ij}(x)
&=\frac{2}{M_{\rm pl}}\int \frac{d^3{\mathbf k}}{(2\pi)^3}
\frac{i}{2k^0}
\mathcal{O}_{ijlm}(k)
\mathcal{T}_{\rm m}^{lm}(k)e^{ikx}
\label{h_sol}
\nn
&\quad +{\rm c.c.}
\,, \\
\varphi_{\mu\nu}(x)
&=\frac{2}{M_G}\int \frac{d^3{\mathbf p}}{(2\pi)^3}
\frac{i}{2p^0}
\mathcal{J}_{\mu\nu}(p)e^{ipx}
+{\rm c.c.} 
\,,
\label{phi_sol}
\end{align}
where 
\begin{align}
\mathcal{O}_{ijlm}=P_{l(i}P_{j)m}-\frac{1}{2}P_{ij}P_{lm}
,
P_{ij}=\delta_{ij}-k_i k_j/{\mathbf k}^2
,
\end{align}
is the transverse-traceless projection operator which is introduced to satisfy the transverse-traceless gauge. Note that the source terms 
\begin{align}
\mathcal{T}_{\rm m}^{ij}(k)&=\int d^4y e^{-iky}T_{\rm m}^{ij}(y)
\,,\\
\mathcal{J}^{\alpha\beta}(p)&=\int d^4y e^{-ipy}J^{\alpha\beta}(y)
\nn
&=\mathcal{T}_{\rm m}^{\alpha\beta}(p)-\frac{1}{3}\left(\eta^{\alpha\beta}+\frac{p^{\alpha}p^{\beta}}{m^2} \right) \mathcal{T}_{\rm m}(p)
\,,
\end{align}
are evaluated at only $k^2=0$ and $p^2=-m^2$, respectively. The on-shell condition for the massive graviton leads $p^{\mu}\mathcal{J}_{\mu\nu}=0,\mathcal{J}^{\mu}{}_{\mu}=0$, thus the massive graviton automatically satisfies the transverse-traceless condition after the source vanishes.
As a result, we find
\begin{widetext}
\begin{align}
\langle h_{\alpha\beta}{}^{,\mu} h^{\alpha\beta,\nu} \rangle 
&=\frac{4}{M_{\rm pl}^2} 
\left\langle
 \int \!\! \frac{d^3 {\mathbf k}}{(2\pi)^3} \int \!\! \frac{d^3 {\mathbf k'}}{(2\pi)^3} 
\frac{k^{\mu} k'{}^{\nu}}{2 k^0 k'{}^0} \mathcal{T}^{kl}_{\rm m}(k) \mathcal{O}^{ij}_{kl}(k) \mathcal{O}_{ijnm}(k')\mathcal{T}^{* nm}_{\rm m}(k')
e^{i(k-k')x} 
\right\rangle
%+(k\leftrightarrow k')
,\\
\langle \varphi_{\alpha\beta}{}^{,\mu} \varphi^{\alpha\beta,\nu} \rangle 
&=\frac{4}{M_G^2}
\left\langle
 \int \!\! \frac{d^3 {\mathbf p}}{(2\pi)^3} 
 \int \!\! \frac{d^3 {\mathbf p}'}{(2\pi)^3} 
\frac{p^{\mu} p'{}^{\nu}}{2 p^0 p'{}^0} \mathcal{J}^{\alpha\beta}(p)\mathcal{J}^*_{\alpha\beta}(p') e^{i(p-p')x}
\right\rangle
%+(p\leftrightarrow p')
\nn
&\approx 
\frac{4}{M_G^2}
\left\langle
 \int \!\! \frac{d^3 {\mathbf p}}{(2\pi)^3} 
 \int \!\! \frac{d^3 {\mathbf p}'}{(2\pi)^3} 
\frac{p^{\mu} p'{}^{\nu}}{2 p^0 p'{}^0} \left( \mathcal{T}_{\rm m}^{\alpha\beta}(p)\mathcal{T}^*_{\rm m}{}_{\alpha\beta}(p')-\frac{1}{3}\mathcal{T}_{\rm m}(p)\mathcal{T}_{\rm m}^*(p')
\right) e^{i(p-p')x}
\right\rangle
%+(p\leftrightarrow p')
, \label{eqn:<varphivarphi>}
\end{align}
\end{widetext}
where $*$ denotes the complex conjugate and we have used the on-shell condition $p^2=-m^2$. While the last term in (\ref{def_J}) would diverge in the limit $m^2\to 0$, (\ref{eqn:<varphivarphi>}) is finite in the same limit. 

The result indicates that, if most of the produced massive gravitons are relativistic, the amount of the gravitons are simply evaluated by
\begin{align}
\langle \varphi_{\alpha\beta}{}^{,\mu} \varphi^{\alpha\beta,\nu} \rangle 
\sim
\frac{M_{\rm pl}^2}{M_G^2} \langle h_{\alpha\beta}{}^{,\mu} h^{\alpha\beta,\nu} \rangle\,.
\label{ratio}
\end{align}
On the other hand, if non-relativistic massive gravitons are generated, the amount of the massive graviton strongly depends on the Fourier space distribution of the source.

\subsection{Non-relativistic production}

First we consider the case where the peak momentum $k_*$ is smaller than the graviton mass, i.e., $m>k_*\sim 1/R_*$, where $R_*$ is the scale of the field bubble. In this case the massive graviton is produced with non-relativistic velocity and continues to be non-relativistic afterwards. Therefore, the massive graviton behaves like a cold dark matter.

In order to relate the abundance of massive gravitons as dark matter to the amount of gravitational waves, we are interested in the ratio of the stress-energy tensors for the massive and massless gravitons. In the case under consideration, i.e. for $m>k_*\sim 1/R_*$, the ratio strongly depends on the value of $mR_*$. Since the bubbly stage of the preheating is significantly non-Gaussian, the estimate of the abundance of massive gravitons requires detailed numerical simulations. We thus consider it beyond the scope of the present paper to discuss further on the case of non-relativistic production.

\subsection{Relativistic production}

Next we consider the case where the peak momentum of the gravitational wave is higher than the graviton mass, i.e. $m<k_*\sim 1/R_*$, where $R_*$ is the scale of the field bubble. In this case the massive graviton is produced with relativistic velocities. In order to realize the bottom-up scenario of the structure formation, we thus need to make it sure that the free streaming scale due to the massive graviton is less than about $0.1$ Mpc~\cite{Viel:2013apy}. The free streaming due to the relativistic motion of massive gravitons continues until the peak momentum is redshifted down to $m$. Therefore, The free streaming scale is estimated as 
\begin{eqnarray}
 L_{\rm fs} &\sim &\frac{a_0}{a_{\rm nr} H_{\rm nr}} 
 \sim \frac{a_{\rm nr}}{a_*}\frac{a_0}{a_*H_*}
 \sim \frac{1}{mR_*}\frac{a_0}{a_*H_*} \nonumber\\
 &\sim &\frac{2\pi f}{m}10^{7} \;{\rm Mpc}\,,
\end{eqnarray}
where $a_{\rm nr}$ and $H_{\rm nr}$ are the scale factor and the Hubble expansion rate, respectively, at the time when the massive graviton becomes non-relativistic and $a_0$ is the scale factor today. By requiring that $L_{\rm fs}$ be less than $0.1$ Mpc, we thus obtain the constraint
\begin{align}
 \frac{m}{2\pi f} > 10^8\,.
\end{align}
Therefore, in the case of the relativistic production, if the characteristic frequency of the gravitational wave from preheating is determined by observation, we can obtain a lower bound on the graviton mass.

\begin{figure*}[htbp]
\centering
\includegraphics[width=15cm,angle=0,clip]{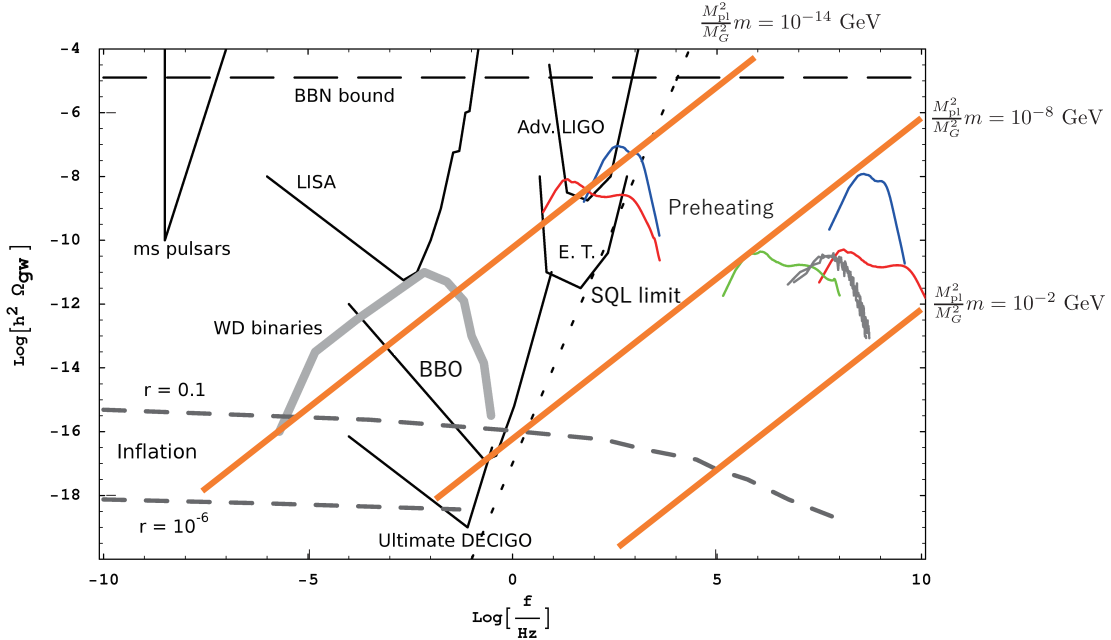}
\caption{The sensitivities of gravitational wave detectors and the expected gravitational wave spectra from the preheating (red, blue, green and gray curves at the right), adopted from \cite{Dufaux:2008dn}. The orange lines then represent expected frequency and amplitude of the gravitational wave background corresponding to the massive graviton dark matter model for $(M_{\rm pl}/M_G)^2\times m=10^{-14}$GeV, $10^{-8}$GeV and $10^{-2}$GeV. The gravitational wave background thus determines the combination $(M_{\rm pl}/M_G)^2\times m$. In particular, some of gravitational wave spectra are detectable by LIGO, for which the massive graviton can be the dominant component of dark matter when $(M_{\rm pl}/M_G)^2\times m\sim 10^{-14}$ GeV.}
\label{fig_sensitivity}
\end{figure*}

In this case, most of the generated massive gravitons are relativistic with the momentum $\sim k_*>m$, thus both massive and massless gravitons are created by the sources with almost the same four-momenta. As shown in \eqref{ratio}, the energy densities are thus evaluated as
\begin{align}
\frac{\rho_G^*}{\rho_{\rm gw}^*}\sim \frac{M_{\rm pl}^2}{M_G^2}\,,
\end{align} 
where $\rho_G^*$ and $\rho_{\rm gw}^*$ are the energy densities of the massive graviton and the massless graviton at the production time. When the massive graviton is relativistic, the energy densities of both gravitons decrease as $a^{-4}$, where $a$ is the scale factor of the Universe. As the Universe expands, the massive graviton becomes non-relativistic, and then the energy density of the massive graviton decreases as $a^{-3}$. Hence the energy density of the massive graviton at the present is
\begin{align} 
\Omega_G
\sim \frac{M_{\rm pl}^2}{M_G^2} \frac{m}{2\pi f}\Omega_{\rm gw}
\,,
\end{align}
where $\Omega_G$ is the density parameter of the massive graviton. Hence if the massive graviton is the dominant component of dark matter, the combination $(M_{\rm pl}/M_G)^2\times m$ can be estimated by the gravitational wave background as shown in Fig. \ref{fig_sensitivity}. 

Since the present abundance and the frequency of the gravitational wave background can be evaluated by \eqref{GW_preheating}, the present abundance and the free streaming scale of the massive graviton can be estimated by using $\rho_*$ and $R_*$. We now focus on gravitational waves to be sensitive in the LIGO range. For instance, the preheating of
\begin{align} 
\rho_*^{1/4}\sim 10^8\, {\rm GeV}\,,\quad
R_*^{-1} \sim 0.1 \,{\rm GeV}\,,
\end{align}
predicts the gravitational wave background with
\begin{align}
f \sim 40 \,{\rm Hz}\,, \quad h^2\Omega_{\rm gw} \sim \alpha\,10^{-8}\,.
\label{LIGO_GW}
\end{align}
Note that the graviton mass has been assumed to be consistent with the Higuchi bound, i.e. $m> \sqrt{2}H_*$, to avoid the Higuchi instability, while the relativistic production is realized only when $m < R_*^{-1}$. Hence, the consistency of our assumptions leads $R_*^{-1} > m>\sqrt{2}H_*$. A set of consistent parameters is 
\begin{align}
m\sim 0.01\; {\rm GeV}\,,\quad
M_G \sim 10^6 M_{\rm pl}\,, 
\label{viable_mass}
\end{align}
in which the massive graviton can explain the observed amount of the dark matter. Since $\sqrt{2}H_* \sim 0.005 {\rm GeV}$, the Higuchi bound is barely satisfied. The corresponding free streaming scale is about $10^{-7}$ pc, so the massive graviton behaves like a cold dark matter. Therefore if the gravitational detectors observe the stochastic gravitational wave background with \eqref{LIGO_GW}, the massive graviton with \eqref{viable_mass} is a viable candidate of dark matter.

%%%%%%%%%%%%%%%%%%%%%%%%%%%%%%%%%%%%%%%%%%%%%%%%%%%
%%%%%%%%%%%%%%%%%%%%%%%%%%%%%%%%%%%%%%%%%%%%%%%%%%%
%%%%%%%%%%%%%%%%%%%%%%%%%%%%%%%%%%%%%%%%%%%%%%%%%%%
%%%%%%%%%%%%%%%%%%%%%%%%%%%%%%%%%%%%%%%%%%%%%%%%%%%
%%%%%%%%%%%%%%%%%%%%%%%%%%%%%%%%%%%%%%%%%%%%%%%%%%%
%%%%%%%%%%%%%%%%%%%%%%%%%%%%%%%%%%%%%%%%%%%%%%%%%%%
%%%%%%%%%%%%%%%%%%%%%%%%%%%%%%%%%%%%%%%%%%%%%%%%%%%
%%%%%%%%%%%%%%%%%%%%%%%%%%%%%%%%%%%%%%%%%%%%%%%%%%%

\section{Concluding remarks}
\label{summary}

We have proposed a scenario in which the massive graviton in the context of the ghost-free bigravity theory is the dark matter in our universe. First, we derived the energy-momentum tensor of the massive graviton from the nonlinear bigravity theory and confirmed that the massive graviton actuary behaves like a dark matter. Then we discussed a generation mechanism and the present abundance of the massive graviton. In this paper, we assumed that the graviton mass is high enough so that the theory is free from the Higuchi instability during and after the generation of massive graviton. Hence we can discuss the cosmological evolution of gravitons by using the linear theory from the generation of the massive gravitons all the way down to the present epoch. 

One implication of our scenario of massive graviton dark matter is that gravitational waves can carry information about the dark matter. While the ghost-freeness of the bigravity theory in the simplest setup requires that matter fields should couple to either $g_{\mu\nu}$ or $f_{\mu\nu}$~\footnote{However, see \cite{DeFelice:2015yha} for the ghost-free double matter coupling in the partially constrained vielbein formulation.}, neither $g_{\mu\nu}$ nor $f_{\mu\nu}$ are mass eigenstates. Instead, $g_{\mu\nu}$ and $f_{\mu\nu}$ are linear combinations of mass eigenstates. For this reason both massless and massive gravitons couple to the same matter fields. As a result, the gravitational wave background and the massive graviton are generated by the same origin. Hence the abundance of the dark matter is related to that of the gravitational wave background. For example, a suitable value of the graviton mass for the dark matter can be estimated by gravitational wave background observations. Furthermore, if the massive graviton is observed directly by another experiment, it gives a consistency relation of the massive graviton dark matter, and then we can identify whether the massive graviton is indeed the dark matter.

Depending on the nature of the production process as well as the value of the graviton mass,  the massive graviton may behave as a hot, warm, or cold dark matter. In the present paper, we assumed a generation of the massive graviton from the preheating after inflation. If LIGO and Virgo detectors observe the gravitational wave background with $f \sim 40$ Hz and $\Omega_{\rm gw} \sim 10^{-8}$, the massive graviton with $m\sim 0.01$ GeV and $M_G \sim 10^6 M_{\rm pl}$ is a viable candidate of dark matter in this scenario. Since the free streaming scale due to the massive graviton in this case is much shorter than kpc, the massive graviton behaves like a cold dark matter as far as the structure formation is concerned.

Although we have focused on stochastic massive gravitons in the present paper, a condensed massive graviton could be a candidate of dark matter as well. For instance, the energy density of the anisotropy of the Bianchi type universe decreases as a dust fluid in the bigravity theory although that in GR decreases as a stiff matter~\cite{Maeda:2013bha}. This fact could be explained by that the anisotropy is a consequence of a condensation of massive gravitons with some direction and the energy density of the non-relativistic massive gravitons decreases as a dust.

In the present paper, in order to avoid the Higuchi instability we have assumed that the Hubble expansion rate at the time of the massive graviton production is lower than the graviton mass. Even with this restriction, we have found that the production of massive graviton from the preheating provides a viable scenario of dark matter in bigravity. If we can relax the assumption of large graviton mass then various other scenarios would become possible, such as the production of massive graviton during inflation. For instance, if we can extend the recently proposed minimal theory of massive gravity~\cite{DeFelice:2015hla,DeFelice:2015moy}, which does not contain scalar and vector degrees freedom in the gravity sector, to the context of bigravity then it would open up many interesting possibilities. For example, we would find a successful scenario of massive graviton dark matter originated from inflation, based on a bigravity version of the minimal theory of massive gravity.

%%%%%%%%%%%%%%%%%%%%%%%%%%%%%%%%%%%%%%%%%%%%%%%%%%%%%%%%%%%%%%%
%%%%%%%%%%%%%%%%%%%%%%%%%%%%%%%%%%%%%%%%%%%%%%%%%%%%%%%%%%%%%%%
\section*{Acknowledgments}
%%%%%%%%%%%%%%%%%%%%%%%%%%%%%%%%%%%%%%%%%%%%%%%%%%%%%%%%%%%%%%%
%%%%%%%%%%%%%%%%%%%%%%%%%%%%%%%%%%%%%%%%%%%%%%%%%%%%%%%%%%%%%%%
K.A. would like to thank Kei-ichi Maeda for useful discussions and comments.  His work was supported in part by Grants-in-Aid from the Scientific Research Fund of the Japan Society for the Promotion of Science  (No. 15J05540). 
The work of S.M. was supported in part by JSPS KAKENHI Grant Number
24540256 and World Premier International Research Center Initiative
(WPI), MEXT, Japan.
%The work of S.M. was supported in part by Grant-in-Aid for Scientific Research 24540256 and the WPI Initiative, MEXT Japan. 

%%%%%%%%%%%%%%%%%%%%%%%%%%%%%%%%%%%%%%%%%%%%%%%%%%
%%%%%%%%%%%%%%%%%%%%%%%%%%%%%%%%%%%%%%%%%%%%%%%%%%
%%%%%%%%%%%%%%%%%%%%%%%%%%%%%%%%%%%%%%%%%%%%%%%%%%
%%%%%%%%%%%%%%%%%%%%%%%%%%%%%%%%%%%%%%%%%%%%%%%%%%
%%%%%%%%%%%%%%%%%%%%%%%%%%%%%%%%%%%%%%%%%%%%%%%%%%

\appendix
\begin{widetext}
\section{Derivation of graviton energy-momentum tensor}
\label{appendix}
In this appendix, we summarize the derivation of energy-momentum tensor of the massive graviton.

In the classical field theory, when the Lagrangian is given,
the canonical energy-momentum tensor can be defined from the Noether's theorem.
However, since the Lagrangian has a freedom to add a total divergence term,
the energy-momentum tensor cannot be defined uniquely.
To remove this ambiguity,
we define the canonical energy-momentum tensor by averaging over a spacetime region. 
Hence
we define the canonical energy-momentum tensor of a symmetric tensor field $\chi_{\mu}$ as
\begin{align}
\Theta_{\chi}^{\mu\nu}:=
\left\langle
-\frac{\delta \mathcal{L}_{\chi}}{\delta (\partial_{\mu}\chi_{\alpha\beta})} \partial^{\nu}\chi_{\alpha\beta}
+\eta^{\mu\nu}\mathcal{L}_{\chi}
\right\rangle\,,
\end{align}
where the symbol $\langle \cdots \rangle$ denotes the average over a spacetime region 
which is assumed to be sufficiently larger than the wave-packet.
For the massless graviton $h_{\mu\nu}$,
the Lagrangian is given by \eqref{EH_quadratic}.
Assuming the transverse-traceless gauge,
the canonical energy-momentum tensor is calculated by
\begin{align}
\Theta_{{\rm gw}}^{\mu\nu}
&=\left\langle
-\frac{\delta \mathcal{L}_{\rm EH}}{\delta (\partial_{\mu}h_{\alpha\beta})} \partial^{\nu}h_{\alpha\beta}
+\eta^{\mu\nu}\mathcal{L}_{\rm EH}
\right\rangle
\nn
&=\frac{1}{4}\langle \partial^{\mu}h^{\alpha\beta}\partial^{\nu}h_{\alpha\beta} \rangle
+\frac{1}{8}\eta^{\mu\nu}\langle h_{\alpha\beta} \partial^2 h^{\alpha\beta} \rangle\,.
\end{align}
The second term vanishes from the field equation \eqref{EH_eom},
and then the canonical energy-momentum tensor of the massless graviton is given by
\begin{align}
\Theta_{\rm gw}^{\mu\nu}=
\frac{1}{4}\langle \partial^{\mu}h^{\alpha\beta}\partial^{\nu}h_{\alpha\beta} \rangle
\,.
\label{canonical_T_massless}
\end{align}
The canonical energy-momentum tensor of the massive graviton can be obtained in a way similar to the case of the massless graviton.
By using the field equation \eqref{FP_eom} and the transverse-traceless condition \eqref{TT_condition},
the canonical energy-momentum tensor is given by
\begin{align}
\Theta_{G}^{\mu\nu }
&=\left\langle
-\frac{\delta (\mathcal{L}_{\rm EH}+\mathcal{L}_{\rm FP})}{\delta (\partial_{\mu}\varphi_{\alpha\beta})} 
\partial^{\nu}\varphi_{\alpha\beta}
+\eta^{\mu\nu}(\mathcal{L}_{\rm EH}+\mathcal{L}_{\rm FP})
\right\rangle
\nn
&=\frac{1}{4}\langle \partial^{\mu}\varphi^{\alpha\beta}\partial^{\nu}\varphi_{\alpha\beta} \rangle\,.
\end{align}

In general relativity, 
the energy-momentum tensor of the graviton defined from Noether's theorem
is also obtained from the nonlinear part of the Einstein equation.
Here we consider up to second order of the perturbation.
For a transverse-traceless perturbation 
$\chi_{\mu\nu}:=M_{\rm pl}(g_{\mu\nu}-\eta_{\mu\nu})$,
the second order part of the Ricci tensor is given by
\begin{align}
M_{\rm pl}^2\delta \overset{  (2)}{R}{}_{\mu\nu}
&=
\frac{1}{4}\chi^{\alpha\beta}{}_{,\mu}\chi_{\alpha\beta,\nu}
-\frac{1}{2}\chi_{\mu\alpha,\beta}\chi_{\nu}{}^{\beta,\alpha}
+\frac{1}{2}\chi_{\mu}{}^{\alpha,\beta}\chi_{\nu\alpha,\beta}
+\frac{1}{2}\chi^{\alpha\beta}(\chi_{\alpha\beta,\mu\nu}
-2\chi_{\alpha(\mu,\nu)\beta}+\chi_{\mu\nu,\alpha\beta})
\,,
\end{align}
where we have imposed the transverse-traceless gauge condition and $\langle\cdots\rangle$ represents an average over a spacetime region.  We define the energy-momentum tensor of the graviton as
\begin{align}
 T^{\mu\nu}_{\chi}:=-
 \left( \eta^{\mu\alpha}\eta^{\nu\beta}
 - \frac{1}{2}\eta^{\mu\nu}\eta^{\alpha\beta} \right)
 M_{\rm pl}^2
 \langle  \delta \overset{  (2)}{R}{}_{\alpha\beta}(\chi) \rangle
\,.
\end{align}
Integrating by part (under the high-frequency/momentum approximation) and using the equation of motion $\chi_{\alpha\beta,\gamma}{}^{,\gamma}=0$, one can obtain 
\begin{align}
T^{\mu\nu}_{\chi}=\frac{1}{4}\langle \chi^{\alpha\beta,\mu} \chi_{\alpha\beta}{}^{,\nu} \rangle\,,
\end{align}
which is the same as the result from the Noether's theorem.
Including the energy-momentum tensors of the graviton as well as the matter,
the Einstein equation is expressed as
\begin{align}
\mathcal{E}^{\mu\nu,\alpha\beta}\chi_{\alpha\beta}
=\frac{1}{M_{\rm pl}} 
(T^{\mu\nu}_{\rm m}+T^{\mu\nu}_\chi)
\,.
\end{align}
Hence the energy-momentum tensor of the graviton is a source of the gravitational field.
Note that the conservation law of the energy-momentum tensor is guaranteed without taking an average over a spacetime region. The divergence of $T^{\mu\nu}_{\chi}$ is calculated as
\begin{align}
\partial_{\nu}T^{\mu\nu}_{\chi}
=\left(\frac{1}{4}\chi^{\alpha\beta,\mu}-\frac{1}{2}\chi^{\mu\alpha,\beta}
\right)\chi_{\alpha\beta,\gamma}{}^{,\gamma}
\,,
\end{align}
which is zero due to the field equation.

The canonical energy-momentum tensor of the massive graviton
would be a source of the gravitational field in bigravity.
We denote the fully nonlinear Einstein equations as
\begin{align}
G^{\mu\nu}(g)&=\kappa_g^2\left( T^{({\rm int})\mu\nu}_g+T^{\mu\nu}_{\rm m} \right)\,,\\
G^{\mu\nu}(f)&=\kappa_f^2T^{({\rm int})\mu\nu}_f\,,
\end{align}
where $T^{({\rm int} ) \mu\nu}_g$ and $T^{({\rm int} ) \mu\nu}_f$ are obtained from the variation of the potential \eqref{dRGT_potential} 
with respect to $g_{\mu\nu}$ and $f_{\mu\nu}$, respectively.
We expand the equations around the Minkowski vacuum up to the second order of perturbations \eqref{perturbations}.
We use the transverse-traceless gauge for the massless graviton continuously.
The second order parts of $T^{({\rm int} ) \mu\nu}_g$ and $T^{({\rm int} ) \mu\nu}_f$
in terms of the mass eigenstates
are given by
\begin{align}
\delta \overset{  (2)}{T}{}_g^{({\rm int} ) \mu\nu} 
&=\frac{m^2}{8\kappa^2}
\left[ (9\kappa_g^2+\kappa_f^2+2c_3 \kappa^2) 
 \varphi^{\mu\alpha}\varphi^{\nu}{}_{\alpha} 
-(4\kappa_g^2+c_3\kappa^2) \varphi^{\alpha\beta}\varphi_{\alpha\beta} \eta^{\mu\nu}
\right]
\nn
&+m^2\frac{\kappa_g\kappa_f}{\kappa^2}
\Biggl[
 \frac{1}{4}\varphi^{\alpha(\mu}h^{\nu)}{}_{\alpha}
-\frac{1}{2}h^{\alpha\beta}\varphi_{\alpha\beta}\eta^{\mu\nu}
\Biggl]
\,,\\
\delta \overset{  (2)}{T}{}_f^{({\rm int} ) \mu\nu}
&=\frac{m^2}{8\kappa^2}
\left[
(-5\kappa_g^2+3\kappa_f^2-2c_3\kappa^2) \varphi^{\mu\alpha}\varphi^{\nu}{}_{\alpha}
-(-3\kappa_g^2+\kappa_f^2-c_3\kappa^2)\varphi^{\alpha\beta}\varphi_{\alpha\beta}\eta^{\mu\nu} \right]
\nn
&+m^2\frac{\kappa_g\kappa_f}{\kappa^2}\left[
-\frac{1}{4}\varphi^{\alpha(\mu}h^{\nu)}{}_{\alpha}
+\frac{1}{2}h^{\alpha\beta}\varphi_{\alpha\beta}\eta^{\mu\nu}
\right]
\,,
\end{align}
where we use $h^{\mu}{}_{\mu}=0$ and $\varphi^{\mu}{}_{\mu}=0$.
Note that although both $T^{({\rm int} ) \mu\nu}_g$ and $T^{({\rm int} ) \mu\nu}_f$ are complicated,
the sum is simply given by
\begin{align}
\delta \overset{  (2)}{T}{}_g^{({\rm int} ) \mu\nu} 
+\delta \overset{  (2)}{T}{}_f^{({\rm int} ) \mu\nu}
=\frac{m^2}{2}\varphi^{\mu\alpha}\varphi^{\nu}{}_{\alpha} 
-\frac{m^2}{8}\varphi^{\alpha\beta}\varphi_{\alpha\beta} \eta^{\mu\nu}
\,.
\end{align}

We find the canonical energy-momentum tensors of massless and massive gravitons
are obtained as source terms of the field equation of the massless graviton.
Including the energy-momentum tensors,
the equation of motion of the massless graviton is expressed by
\begin{align}
\mathcal{E}^{\mu\nu,\alpha\beta}h_{\alpha\beta}
=\frac{1}{M_{\rm pl}}(T_{\rm m}^{\mu\nu}+T_{\rm gw}^{\mu\nu}+T_G^{\mu\nu})
\,,
\end{align}
where
the energy-momentum tensors 
of the massless graviton and the massive graviton are defined by
\begin{align}
T^{\mu\nu}_{\rm gw}&:=-M_{\rm pl}^2 \delta \overset{(2)}{G}{}^{\mu\nu}(h)
\nn
&\:=
-\frac{1}{4}h^{\alpha\beta,\mu}h_{\alpha\beta}{}^{,\nu}
+\frac{1}{2}h^{\mu}{}_{\alpha,\beta}h^{\nu\beta,\alpha}
-\frac{1}{2}h^{\mu\alpha,\beta}h^{\nu}{}_{\alpha,\beta}
\nn
&\:
-h^{\alpha(\mu}h^{\nu)}{}_{\alpha,\beta}{}^{,\beta}
-\frac{1}{2}h^{\alpha\beta}
(h_{\alpha\beta}{}^{,\mu\nu}
-2h^{(\mu}{}_{\alpha}{}^{,\nu)}{}_{\beta}+h^{\mu\nu}{}_{,\alpha\beta})
\nn
&\:
+\eta^{\mu\nu}
\left(\frac{3}{8}h_{\alpha\beta,\gamma}h^{\alpha\beta,\gamma}
-\frac{1}{4}h_{\alpha\beta,\gamma}h^{\alpha\gamma,\beta}
+\frac{1}{2}h^{\alpha\beta}h_{\alpha\beta,\gamma}{}^{,\gamma}\right)
\,, \\
T^{\mu\nu}_G&:=
-\frac{1}{\kappa^2} \delta \overset{  (2)}{G}{}^{\mu\nu}(\varphi)
+\delta \overset{  (2)}{T}{}^{({\rm int} ) \mu\nu}_g+
\delta \overset{  (2)}{T}{}^{({\rm int} ) \mu\nu}_f
\nn
&\:
=
-\frac{1}{4}\varphi^{\alpha\beta,\mu}\varphi_{\alpha\beta}{}^{,\nu}
+\frac{1}{2}\varphi^{\mu}{}_{\alpha,\beta}\varphi^{\nu\beta,\alpha}
-\frac{1}{2}\varphi^{\mu\alpha,\beta}\varphi^{\nu}{}_{\alpha,\beta}
\nn
&\:
-\varphi^{\alpha(\mu}\varphi^{\nu)}{}_{\alpha,\beta}{}^{,\beta}
-\frac{1}{2}\varphi^{\alpha\beta}(\varphi_{\alpha\beta}{}^{,\mu\nu}
-2\varphi^{(\mu}{}_{\alpha}{}^{,\nu)}{}_{\beta}+\varphi^{\mu\nu}{}_{,\alpha\beta})
\nn
&\:
+\eta^{\mu\nu}
\left(\frac{3}{8}\varphi_{\alpha\beta,\gamma}\varphi^{\alpha\beta,\gamma}
-\frac{1}{4}\varphi_{\alpha\beta,\gamma}\varphi^{\alpha\gamma,\beta}
+\frac{1}{2}\varphi^{\alpha\beta}\varphi_{\alpha\beta,\gamma}{}^{,\gamma}\right)
\nn
&\:
+\frac{m^2}{2}\varphi^{\mu\alpha}\varphi^{\nu}{}_{\alpha} 
-\frac{m^2}{8}\varphi^{\alpha\beta}\varphi_{\alpha\beta} \eta^{\mu\nu}
\,.
\end{align}
Averaging over a spacetime region, 
the energy-momentum tensors are reduced into
\begin{align}
T^{\mu\nu}_{\rm gw}&=\frac{1}{4}\langle h^{\alpha\beta,\mu} h_{\alpha\beta}{}^{,\nu} \rangle\,,
\\
T^{\mu\nu}_G&=\frac{1}{4}\langle \varphi^{\alpha\beta,\mu} \varphi_{\alpha\beta}{}^{,\nu} \rangle\,,
\end{align}
which are indeed the same as the canonical energy-momentum tensors defined from Noether's theorem.
The energy-momentum tensors $T^{\mu\nu}_{\rm gw}$ and $T^{\mu\nu}_G$ satisfy the conservation laws without the average over a spacetime region. 
The energy-momentum tensor of the massless graviton is the same as that in the case of GR, while the divergence of that of the massive graviton is calculated by
\begin{align}
\partial_{\nu}T^{\mu\nu}_G
=\left(\frac{1}{4}\varphi^{\alpha\beta,\mu}-\frac{1}{2}\varphi^{\mu\alpha,\beta}\right)
(\varphi_{\alpha\beta,\gamma}{}^{,\gamma}-m^2\varphi_{\alpha\beta}) 
\,.
\end{align}
Hence the conservation law of the energy-momentum tensor of the massive graviton is guaranteed as well.
As a result,
we conclude that, in bigravity,
both massless and massive gravitons are sources of 
the gravity mediated by the massless graviton rather than either $g_{\mu\nu},f_{\mu\nu}$ or the massive graviton.

\end{widetext}

%%%%%%%%%%%%%%%%%%%%%%%%%%%%%%%%%%%%%%%%%%%%%%%%%%%%%%%%
%%%%%%%%%%%%%%%%%%%%%%%%%%%%%%%%%%%%%%%%%%%%%%%%%%%%%%%%
%%%%%%%%%%%%%%%%%%%%%%%%%%%%%%%%%%%%%%%%%%%%%%%%%%%%%%%%
%%%%%%%%%%%%%%%%%%%%%%%%%%%%%%%%%%%%%%%%%%%%%%%%%%%%%%%%
%%%%%%%%%%%%%%%%%%%%%%%%%%%%%%%%%%%%%%%%%%%%%%%%%%%%%%%%
%%%%%%%%%%%%%%%%%%%%%%%%%%%%%%%%%%%%%%%%%%%%%%%%%%%%%%%%
%%%%%%%%%%%%%%%%%%%%%%%%%%%%%%%%%%%%%%%%%%%%%%%%%%%%%%%%

%%%%%%%%%%%%%%%%%%%%%%%%%%%%%%%%%%%%%%%%%%%%%%%%%%%
%%%%%%%%%%%%%%%%%%%%%%%%%%%%%%%%%%%%%%%%%%%%%%%%%%%
%%%%%%%%%%%%%%%%%%%%%%%%%%%%%%%%%%%%%%%%%%%%%%%%%%%
%%%%%%%%%%%%%%%%%%%%%%%%%%%%%%%%%%%%%%%%%%%%%%%%%%%
%%%%%%%%%%%%%%%%%%%%%%%%%%%%%%%%%%%%%%%%%%%%%%%%%%%
%%%%%%%%%%%%%%%%%%%%%%%%%%%%%%%%%%%%%%%%%%%%%%%%%%%
%%%%%%%%%%%%%%%%%%%%%%%%%%%%%%%%%%%%%%%%%%%%%%%%%%%
%%%%%%%%%%%%%%%%%%%%%%%%%%%%%%%%%%%%%%%%%%%%%%%%%%%

\bibliographystyle{apsrev4-1}
\bibliography{ref}

\end{document}